\documentclass[a4paper,11pt]{article}
\pdfoutput=1 

\usepackage{jinstpub} 

\usepackage{lineno}
\usepackage{subfig}
\usepackage{float}

\title{\boldmath Nuclearite search with ANTARES}

\author[a, 1]{M. Bouta, \note{Corresponding author.}}
\author[b]{J. Brunner}
\author[a]{A. Moussa}
\author[c]{G. E. P\u{a}v\u{a}la\c{s}}
\author[d]{Y. Tayalati}

\affiliation[a]{University Mohammed First in Oujda,\\
  67600, Oujda, Morocco}
\affiliation[b]{Centre de Physique des Particules de Marseille, France}
\affiliation[c]{Institute of Space Science, M\u{a}gurele, Romania}
\affiliation[d]{University Mohammed V in Rabat, Morocco}

\emailAdd{boutamohammed@hotmail.com}

\abstract{
The ANTARES detector is a Cherenkov underwater neutrino telescope operating in the Mediterranean Sea. Its construction was completed in 2008. Even though optimised for the search of cosmic neutrinos, this telescope is also sensitive to nuclearites (massive nuggets of strange quark matter) trough the black body radiation emitted along their path.

We discuss here the possible detection of non-relativistic down-going nuclearites with the ANTARES telescope and present the results of an analysis using data collected from 2009 till 2017.
} 


\collaboration[c]{on behalf of ANTARES Collaboration}

\proceeding{9$^{\text{th}}$ Workshop on Very Large Volume Neutrino Telescope \\
  18 - 21 May 2021\\
  Valencia}

\usepackage{setspace}
\AtBeginDocument{%
  \addtolength\abovedisplayskip{-0.5\baselineskip}%
  \addtolength\belowdisplayskip{-0.5\baselineskip}%
}

\begin{document}
\maketitle
\flushbottom

\section{Introduction}
\label{sec:intro}
Nuclearites are hypothetical heavy particles derived from the strange quark matter (SQM) Theory \cite{EWitten}. They could be present in the cosmic radiation reaching the Earth originating from relics of the early Universe as nuggets or strange star collisions. They could be detected by using the light generated by their atomic collision as they cross a transparent mediums \cite{DeRujula&Glashow,Popa}.

In this work, we study cosmic nuclearites falling on the Earth with galactic velocities $\beta = 10^{-3}$, assuming a mass of $4 \times 10^{13} $  GeV/c$^2$ as a threshold mass for nuclearites detection at the ANATRES level. By using a dedicated Monte Carlo simulation, each particle is generated with random zenith and azimuth angles and its velocity is evaluated, then propagated throw the detector where the signals are produced. A first results on sensitivities of ANTARES to these particles using data collected in the period 2009-2017 is presented.

\section{The ANTARES detector}
\label{sec:Antares}

ANTARES (Astronomy with a Neutrino Telescope and Abyss environmental RESearch) is a Cherenkov based neutrino telescope, deployed at a depth of 2450 m in the Mediterranean Sea, 42 km offshore from Toulon in France. The detector, completed in 2008, is made of 12 vertical detection lines 450 m long, horizontaly spaced by about 60-75 m. Each line contains 25 floors separated by 14.5 m while the first one starts 100 m from the bottom of the sea. Each floor consists of three detection units called optical modules (OMs) comprising a 10" photomultiplier tube (PMT). Each line is anchored to the seabed with the bottom string socket and a dead weight, and is held vertical by a buoy at the top. The detector covers a surface area of 0.1 km$^2$. When neutrinos interact in the vicinity of the detector, they produce relativistic charged particles which emit Cherenkov light that is detected by the PMTs. A full description of the ANTARES detector can be found in \cite{Aguilar}.

In ANTARES, a "hit" is defined as a  couple of charge and time information. L0 threshold hits are defined as hits satisfying the threshold condition (typically set to 0.3 photo-electrons), and L1 hits are defined as, either a coincidence of two L0 hits from the same storey within 20 ns or a single hit with high amplitude exceeding a predefined High-Threshold condition (set to 3 p.e. or 10 p.e. depending on the data acquisition conditions).

\section{Energy loss of nuclearites}
\label{sec:ELN}
Nuclearites loss their energy mainly by atomic collisions while they cross the matter. For massive nuclearites the energy loss rate is given by \cite{DeRujula&Glashow},

\begin{equation}
-\dfrac{dE}{dx} = \rho \sigma v^2
\end{equation}

\begin{equation}
\sigma(cm^2) =
 \begin{cases} 
  \pi \times 10^{-16} & if\quad  M_{N} < 8.4 \times 10^{14} \ \text{ GeV/c}^2  \\ 
  \pi \times (\dfrac{3M_{N}}{4 \pi \rho_{N}}) & if\quad  M_{N} \geq 8.4 \times 10^{14} \text{ GeV/c}^2 
  \end{cases} 
\end{equation}

\noindent
, Where $\rho$ is the density of the traversed medium, $\sigma$ the effective nuclearite interaction cross-section and $v$ its velocity. $M_{N}$ is the nuclearite mass and $\rho_{N} = 3.5 \times 10^{14} \ g/cm^3$ its density. The cross-section can be obtained by the nuclearite density $\rho_{N}$. 

For nuclearite of mass $M_N$ and an initial velocity of $v_{0}$ penetrating at depth of $L$ in a medium of density $\rho$, its velocity changes as \cite{DeRujula&Glashow,EFarhi&RLJaffe} :

\begin{equation}
v(L) = v_{0} \times \exp \left( {\dfrac{\sigma}{M_N} \int_0^L \rho(x) dx } \right)
\end{equation}

In the case of a transparent medium, nuclearites signal could be tracked by using their visible light emission as a black-body radiation from an expanding cylindrical thermal shock wave \cite{DeRujula&Glashow}. The fraction of energy transformed in light, called luminous efficiency $\eta$, it was estimated to be $\eta \simeq 3 \times 10^{-5}$ in water. Thus, the number of visible photons radiated per unit of path length is estimated as \cite{Gabriela} :

\begin{align}
\dfrac{dN_{\gamma}}{dx} = \dfrac{\eta}{\pi(\text{eV})} \cdot \dfrac{dE}{dx} 
\label{equ:light}
\end{align}

\if false
At the ANTARES level and under the previous condiction the photomultipliers (PMTs) output signal was estimated to be\cite{Gabriela} 

\begin{align}
I \propto \dfrac{\Omega}{4\pi} \times N_{\gamma} \times \exp \bigg(\dfrac{-r}{\lambda_{att}}\bigg)
\label{equ:pmtResponce}
\end{align}

$N_{\gamma}$ is the number of visible photons isotropically emitted along the nuclearite path, $r$ is the distance from the nuclearite position to the PMT in question, and $\lambda_{att}$ is the light attenuation length in sea water, $\lambda_{att} \simeq 50 m$.

$\Omega$ is the solid angle from which the Optical module is seen from the emission point:

\begin{align}
\Omega = \dfrac{ A_{eff} \times \cos(\theta) }{r^{2}}
\end{align}

Where $\theta $ is the incidence angle and $A_{eff}$ is the PMT tube effective area.
\fi

\section{Nuclearites search with ANTARES}
\label{sec:NSA}

To avoid any biased results,  a fraction of experimental data were used to compare real data and the Monte Carlo simulation. Due to the environmental optical background wich affects our analysis, a strong selection criteria is applied for the runs samples in order to select a clean data set.
A Monte Carlo simulation of nuclearite in ANTARES is performed using a dedicated software, a FORTRAN based program that generates these particles with random zenith and azimuth angles on an abstract hemisphere of 548 m radious surrounding the detector.
The detector response was simulated using a dedicated software provided by the collaboration.
The 3D trigger checks for hits causally connected along the same trajectory, it requires at least 5 L1 hits (see section \ref{sec:Antares}) within 2.2 $\mu$s of each other and the T3 trigger is defined as the occurrence of at least two L1 hits in three consecutive storeys within a coincidence time window. This coincidence time window is 100 ns in the case that the two storeys are adjacent, and 200 ns in the case of next to adjacent storeys.

\begin{figure}[htbp]
     \centering
     \includegraphics[width=.49\textwidth, height=5cm]{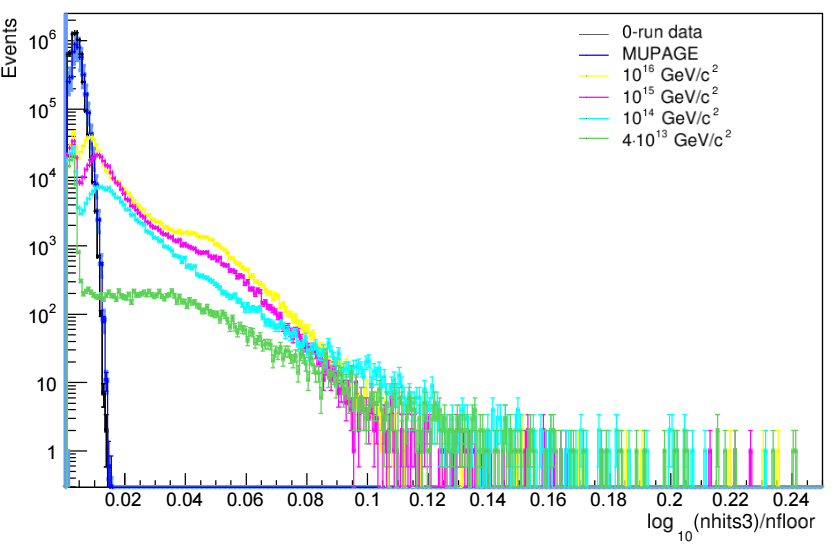}
     \includegraphics[width=.49\textwidth, height=5cm]{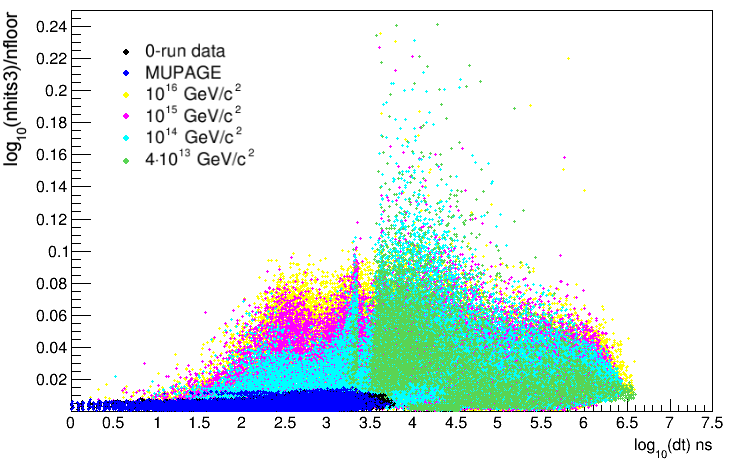}

     \caption{On the left, distribution of log$_{10}$(nhits3)/nfloor; on the right, 2D distribution of log$_{10}$(nhits3)/nfloor versus $dt$. In black, a sample of real data, in blue atmospheric muons, and for different nuclearites masses (other colors).}
     \label{fig:DiscriminationVariables}
\end{figure}

 In this study, we select only events Triggered by the 3D and/or the T3 conditions. A quality cut requiring events with at least 300 L0 hits is applied. This cut allows to reduce the optical noise effect presented in the real data, it allows also to remove the atmospheric muons events of high amplitude and low number of floors.
With the aim of isolating nuclearites signals from the background, two discrimination variables are used. The first one denoted as log$_{10}$(nhits3)/nfloor and shown in figure \ref{fig:DiscriminationVariables},  is the combination of the nhits3 variable which is the number of hits with charge of at least  3 p.e. for a given event and the nfloor variable which represents the number of floors which recorded the hits.  The second discrimination variable, $dt$, refers to the time spent by an event in the detector.  Nuclearites, as they are slow heavy particles, they are expected to have a high transit time in the detector (as shown in the figure \ref{fig:DiscriminationVariables} right).

So as to achieve the best sensitivity of ANTARES to nuclearites without any biased results, the Model Rejection Factor (MRF) is used \cite{Hill}. By using a dedicated algorithm, we look for the best cuts on $dt$ and log$_{10}$(nhits3)/nfloor that minimize the MRF for each nuclearites mass.

\begin{figure}[H]
     \centering
     \includegraphics[width=.49\textwidth, height=5cm]{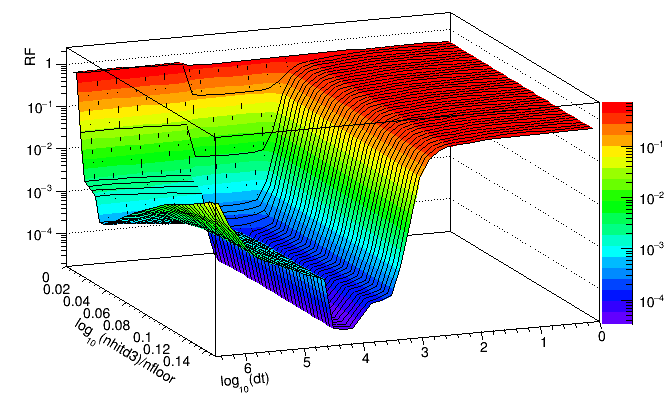}\label{two}
     \includegraphics[width=.49\textwidth, height=5cm]{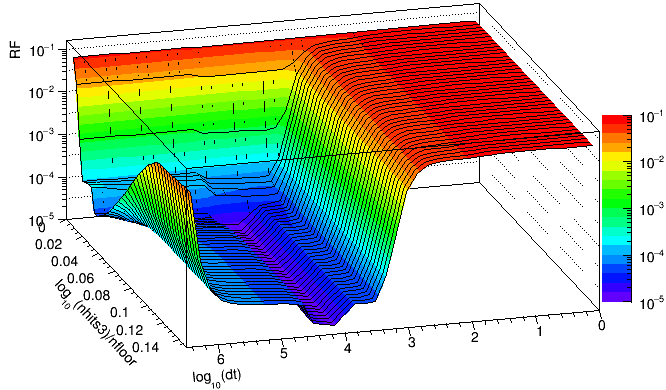}\label{two}
    
     \caption{Examples of the MRF as function of log$_{10}$(nhits3)/nfloor and $dt$ for the lower mass nuclearite $4 \times 10^{13}$ GeV/c$^2$ (left) and the higher mass nuclearite $10^{16}$ GeV/c$^2$ (right).}
     \label{fig:MRF}
\end{figure}

The sensitivity at 90\% confidence level (C.L.) noted $S_{90}$ is computed using the Feldman-Cousins formula \cite{FeldmanCousins} assuming events with a Poissonian distribution.

\begin{equation}
S_{90} (cm^{-2} \cdot sr^{-1} \cdot s^{-1}) =\dfrac{\bar{\mu}_{90}(n_{b})}{S_{eff} (cm^{2} \cdot sr) \times T(s)}
\end{equation}

\begin{equation}
\bar{\mu}_{90} =  \sum_{n_{obs}=1}^{\infty} \mu_{90}(n_{obs}, n_{b}) \times \dfrac{n_{b}^{n_{obs}}}{n_{obs} !} \times e^{-n_{b}}
\end{equation}

\begin{equation}
S_{eff} = \dfrac{n_{Nuc}}{\Phi_{Nuc}}
\end{equation}

\noindent
, where T is the duration of data taking corresponding to the 2009-2017 period in seconds, $n_{Nuc}$ represents the number of nuclearites remaining after applying the optimized cuts, and $\Phi_{Nuc}$ represents the flux of generated Nuclearites. The model Rejection Factor technique consists of varying the cuts with small steps until the minimum of MRF is reached, which corresponds to the values that give the best sensitivity. Figure \ref{fig:MRF} shows an example of the MRF for the lower and the higher nuclearites masses.

\section{Results and conclusion}
\label{sec:result&conclusion}

\begin{figure}[htbp]
     \centering
     \includegraphics[width=.65\textwidth, height=6.0cm]{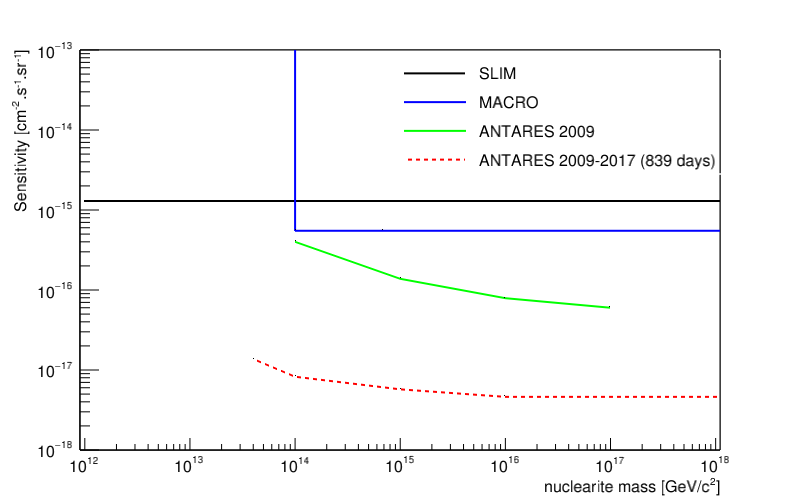}\label{two}
    
     \caption{Sensitivity of ANTARES to a down-going flux of nuclearites (red dotted line), using nine years of data taking. The green line corresopnds to a previous ANTARES published result with less statistics \cite{Pavalas}. Results from other experiments are also shown \cite{MACRO, SLIM}.}
     \label{fig:sens}
\end{figure}

The sensitivity obtained by ANTARES for the period 2009-2017 corresponding to 839 days of livetime of the detector is shown in Figure \ref{fig:sens}. ANTARES  is sensitive to slow moving heavy particles such as nuclearites. For masses higher than $10^{16}$  GeV/c$^{2}$, nuclearites events must be more energetic and they would emit more light. Therefore, the limit of the last test point can be taken as a conservative limit also for larger nuclearites masses.

The result of the sensitivity obtained is competitive to the limits on the flux reported by other experiments such as MACRO and SLIM \cite{MACRO, SLIM}. It also improves the results already obtained by ANTARES for the 2009 period \cite{Pavalas}.

In this work, we presented the analysis performed to search for down-going heavy and slow moving nuclearites assuming a velocity at the entrance of the atmosphere of $\beta_{0}=10^{-3}$. Nuclearites heavier than $4 \times 10^{13}$  GeV/c$^2$ would be able to reach the ANATRES detector depth with enough energies to generate a sufficient visible photons and to be detected. Our study is based on two discrimination variables that reflect perfectly the nuclearite behaviour in the telescope, a run-by-run version 4 Monte Carlo simulations were used where atmospheric muons and optical background such as bioluminescence and $^{40}$K decay present a background for the analysis. The results obtained show that the ANTARES telescope is sensitive to slow moving particules such as nuclearites even in the absence of a dedicated trigger. If no candidate found in the forthcoming analyses, the new flux limit will improve the MACRO and SLIM upper limits.

\end{document}